\documentclass[fleqn,10pt]{wlscirep}
\usepackage[utf8]{inputenc}
\usepackage[T1]{fontenc}
\usepackage{bbm}

\newcommand{\ket}[1]{\left\vert#1\right\rangle}
\newcommand{\bra}[1]{\left\langle#1\right\vert}

\newcommand{\abs}[1]{\left|#1\right|}

\usepackage{physics}
\def\bra#1{\langle #1|}
\def\ket#1{\left|#1 \right>}

\def\Tr{\mbox{Tr}}

\usepackage[normalem]{ulem}
\title{Gravity mediated entanglement between light beams as a table-top test of quantum gravity}

\author[1,*]{Stefan Aimet}
\author[1]{Hadrien Chevalier}
\author[1]{M. S. Kim}
\affil[1]{QOLS, Blackett Laboratory, Imperial College London, London SW7 2AZ, United Kingdom}

\affil[*]{stefan.aimet@gmail.com}

\begin{abstract}
Over the past century, a large community within theoretical physics has been seeking a unified framework for quantum gravity. Yet, to date, there is still no experimental evidence of any non-classical features of gravity.  While traditional experimental proposals would usually require immensely challenging Planck scale experiments, recent table-top protocols based on low-energy quantum control have opened a new avenue into the investigation of non-classical gravity. 
An approach that has sparked high interest, both in terms of experimental feasibility and of theoretical implications, is the indirect witnessing of non-classical gravity through the detection of its capacity to act as an entangling channel. Most discussions have been centred on the entanglement generation between two gravitationally coupled massive systems. In this work, we instead examine the entangling capacity of the gravitational interaction between two light pulses. We explain the main experimental and theoretical advantages of having a photonic protocol, and lay out the steps leading to the determination of the entangling phase, using the path integral formalism and linearised gravity. We establish a closed form formula for the entangling phase and provide an estimated order of magnitude of the average photon number required for the generation of appreciable phase. Using statistical analysis, we show how entanglement may be certified with lower phase signal.
\end{abstract}

\begin{document}
\flushbottom
\maketitle

\thispagestyle{empty}

\section*{Introduction}
 The two main pillars of theoretical physics, general relativity and quantum theory, have been remarkably successful. Experiments have agreed with the predictions of the two theories in their respective domains to very high accuracy, but it is still an open question how exactly to incorporate the two theories into a common framework, which is vaguely labelled as quantum gravity. The hope is to give a more comprehensive account of gravity valid at very high energies or small distances~\cite{oriti_2009}. Often general relativity is described as the theory of the very big, and quantum theory as the theory of the very small scales. Yet, this is not necessarily true, and there are situations where understanding both frameworks together is paramount. This would include better insight into early universe cosmology and black holes as well as the unification of all interactions~\cite{Goenner:2005xs}. The enterprise of formulating such a theory has been a fertile ground for various models, for example string theory~\cite{polchinski1998,green2012}, loop quantum gravity~\cite{rovelli_2004}, causal set theory~\cite{Surya_2019} and many others.

However, all these approaches towards a theory of quantum gravity have been unguided by experimental progress, as the gravitational fields sourced by systems displaying quantum behaviour is mostly insignificant. Vice versa, quantum deviations to the classical Newtonian or Einsteinian descriptions are completely insignificant in the regimes where gravitational effects have importance. The amplitudes of the gravitational coupling and of known quantum couplings become comparable at the Planck scale. Due to its remoteness from any current and future achievable energy levels, to say there is little hope for generating on-demand Planck scale effects in particle accelerators is an understatement. Instead, the natural arena of Planck scale effects consists of black holes and other naturally occurring high-energy cosmic events. Most conventional proposed experimental tests~\cite{Kiefer2004, Laemmerzahl} of quantum gravity include the evaporation of black holes, quantum gravity corrections to the cosmic microwave background or perhaps tests on the discreteness of space. In recent years, much attention has been given to a completely new approach involving table-top tests of gravity at low energies. At these low energies, many fundamental theories of quantum gravity may share the same phenomenology. This means that table-top tests may not rely on testing a specific model of quantum gravity, but they are often model-agnostic~\cite{Huggett2022}.

The history of table-top test proposals claiming to probe quantum gravitational effects~\cite{Pikovski_2012, Bose_2017, Marletto_2017, Krisnanda_2020, carlesso2019testing, howl2021non, Carney2021, Huggett2022} along with surrounding discussions and debates~\cite{Hall2018_2, marshman2020locality, Rydving2021, Doener_2022, Fragkos_2022, MartinMartinez2022, ma2022limits, hosten2022constraints, bose2022mechanism} is a short but rich one. Most notably, the experimental proposals by Bose et al.~\cite{Bose_2017} and Marletto and Vedral~\cite{Marletto_2017} introduced the idea of gravitationally coupling two massive systems that are both in spatially separated superposition states. The gravity mediated entanglement (GME) is then understood as evidence for the quantum nature of gravity. This implication originated first from the quantum information theory framework by using the fact that local operations and classical communication described by a mediator $\mathcal{F}$ cannot generate entanglement
between any two systems $Q_{1,2}$~\cite{Bennet1999, Horodecki2009}, and was later corroborated by using a more general approach~\cite{Marletto_2020}. While there is dissent about the implications of the original protocol, specifically regarding the conclusions one can draw from entanglement generation~\cite{Fragkos_2022,MartinMartinez2022, Hall2018, Hall2018_2, Rydving2021,Doener_2022}, this experimental proposal using gravitational \textit{cat states} has generated a lot of interest in the community.

As innovative as it is, the original massive GME protocol still suffers from several experimental challenges and limitations, some of which pertain to engineering and experimental capabilities, while others are intrinsic to the protocol. From the point of view of experimental implementation, among the most critical obstacles, we may mention the coherent control of a microdiamond of mass on the order of $10^{-14}\,\text{kg}$, which is several orders of magnitude beyond current cutting-edge quantum control~\cite{fein2019quantum, Wood2022,tebbenjohanns2021quantum}. Furthermore, the mismatch between NV centres and centres of mass comes with an unwanted torque, the absence of which anyhow does not eliminate the issue of free-rotation~\cite{ma2021torque}. Another concern is the establishment of statistics which would require particle recycling and implies potential overheating issues~\cite{Bateman_2014,stickler2016spatio, frangeskou2018pure,pedernales2019motional, bykov2019direct}. Finally, using matter to carry out the experiment comes with the unavoidable presence of the Casimir effect~\cite{Casimir1948}, which can only be dealt with by relying on even greater repeatability. From a theoretical point of view, as argued in Ref.~\cite{Christodoulou2022}, the original experiment with massive objects provides only a playground to test the non-relativistic Newtonian regime of gravity but fails to recognise any potential features of gravity as a relativistic quantum field. This aspect is one of several that led to objections to the claim that the original massive GME protocol can certify the non-classicality of gravity.

To resolve some of the challenges faced by the traditional GME protocol using massive objects, this paper suggests the use of light beams as an alternative physical platform to realise GME. General relativity predicts that radiation, as much as matter, sources a gravitational field, the properties of which have been extensively studied~\cite{Tolman1931, Nackoney1977,Scully1979,Mitskievic1989,Ivanov_2003,Schneiter_2019}. The optical equivalent to the matter Stern-Gerlach operation which performs a spin-dependent spatial splitting would be polarisation-dependent beam-splitting, and analogously to the matter experiment where one measures a spin entanglement witness, we would look at a polarisation entanglement witness.

An obvious advantage of the photonic GME protocol is the lack of any undesirable interactions other than gravity. The effect of the direct, short-range photon-photon scattering~\cite{Sangal_2021} can be safely neglected as long as the beams are not overlapping, and for low-energy photons. Thus, we shift the problem from discriminating between different relevant and competitive sources to discriminating between background and signal. It is not hard to imagine why such a variant may be prohibitively challenging: the phase signal due to the gravitational coupling of light is expected to be extremely minute. However, our analysis serves as a first attempt to quantify \textit{how challenging} this approach may be. As stated previously, even if the entanglement generation may be weaker than in the massive protocol - which by the way also requires unreasonably large masses for appreciable signal- there is certainly a case for such an investigation in light of near-future experimental and technological advances. Modern laser technology~\cite{Silva_2021,Kiani2022} grants tunability and control of high-intensity light beams with an unprecedented capacity for empirical repeatability. As we shall see, efficient generation of large amounts of low-noise data will help alleviate the power requirements. While a single photon may only have a negligibly small effect on gravity mediated entanglement due to its small coupling to gravity, the collective effect and reliability of light as a source of entanglement generation may outscore any massive counterpart as a platform for witnessing GME.

Using light also provides a natural framework for the investigation of both relativistic and quantum effects when testing quantum gravity. Due to the easier tunability of frequency - as opposed to mass - finer features of gravity and other deviations from classicality may be more easily probed. Furthermore, light may be viewed as a more convincing candidate, insofar as gravity can be conclusively inferred to be non-classical when GME experiments are performed within light-crossing times between spatial branches~\cite{MartinMartinez2022}. For longer interaction times, classical fields can mediate entanglement as well without the need for quantised gravitational degrees of freedom. Christodoulou et al~\cite{Christodoulou2022} further showed that a fully local approach towards calculating the gravitational phase using the path integral formalism (i.e. without assuming any instantaneous interactions such as in the Newtonian limit) further strengthens the case for demonstrating the non-classicality of gravity. This manifestly local and gauge invariant way of calculating the induced phases is, of course, the natural arena for considering relativistic light beams.

Incidentally, this work may also be seen as a pathway to demonstrate the gravitational coupling of light beams experimentally for the first time. The experimental verification of gravitational coupling of light beams regardless of entanglement, would by itself be an achievement. Various aspects of detecting the gravitational field of light beams were considered in Ref.~\cite{Spengler2022}. Only recently, the role of light has also begun to become more popular in the research community for relativistic GME. For example, the theoretical possibility of photon-matter entanglement was studied in Ref.~\cite{biswas2022gravitational}. Similarly, the interaction of photons was investigated in Refs.~\cite{ratzel2016effect, Mehdi2022}. In this work, we remain in the original double interferometer setup, and use the path integral formalism for our calculations.

Our work is structured as follows. We begin by laying out the setup and introducing basic notions and approximations. We give a brief overview of the path integral description of the experiment, and how one extracts a phase evolution. We also present some instructive and well-established calculations on the metric perturbations sourced by a single circularly polarised light pulse and introduce further helpful notations. Building on this basic situation, we construct the metric perturbation for two counter-propagating pulses and derive the action for two circularly polarised counter-propagating, spatially separated light pulses. After showing some numerical estimations of the gravitationally generated phase, we undertake a statistical analysis of entanglement verification, which drastically lowers the required beam power. We discuss some further directions worth examining, such as improvements, further challenges, and more sophisticated models to describe the photonic GME protocol.

\section*{Results}
\subsection*{Setup}

\begin{figure}[h]
    \centering
    \includegraphics[scale=0.55]{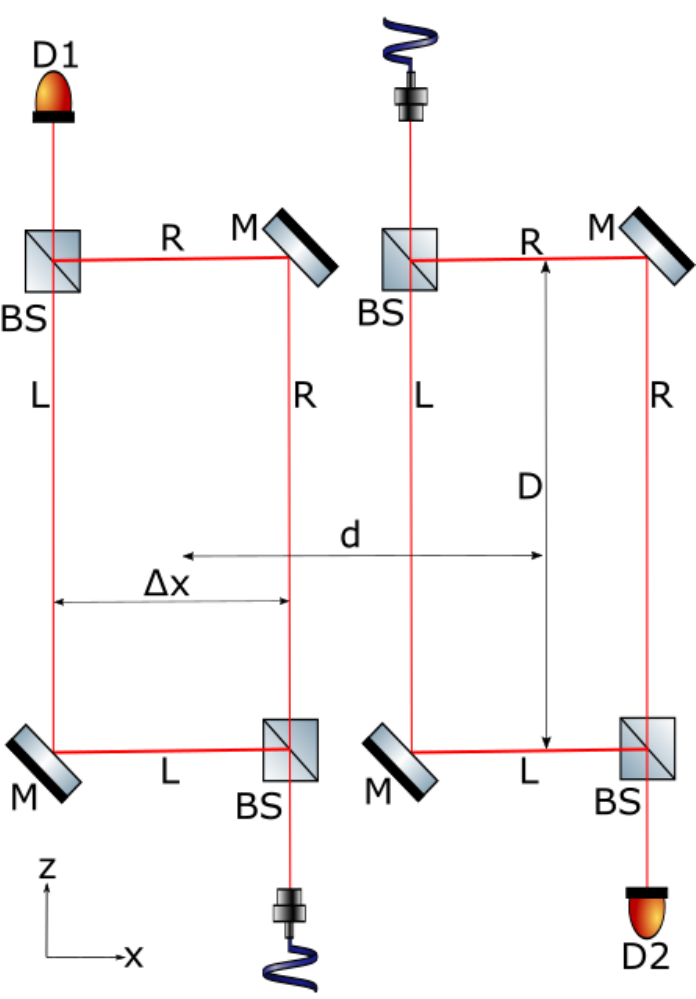}
    \caption{The Mach-Zehnder interferometer setup of the photonic GME protocol. Two counter-propagating light pulses of length $L$ traverse a polarising beam splitter upon their entry into the interferometer of length $D$ which, for certain states, entangles their polarisation degree of freedom with their spatial path. With the potential insertion of quarter waveplates each beam splits into left- and right-handed circular polarisation. The two light pulses interact gravitationally. At the end of the protocol, each spatial branch coincides again meeting at the second polarising beam splitter that disentangles, after the use of waveplates, the polarisation from the spatial mode. The entanglement is then witnessed by local polarisation measurements.}
    \label{fig:mz interferometer}
\end{figure}

Let us consider two Mach-Zehnder interferometers, as shown in Fig.~\ref{fig:mz interferometer}, that receive two counter-propagating light pulses as inputs. The length of each interferometer is $D$ and the transverse separation of the two centres is $d$. Each input pulse enters through a polarising beam splitter. The reason why we chose two counter-propagating light pulses is that co-propagating non-overlapping pulses in vacuum do not interact~\cite{Raetzel_2016,Scully1979}.\\

In general, the GME protocol consists of systems $Q_{a}$ (where $a=1,2$), that are both described by their spatial motion $\vec{x}_a^{s_a}(t)$, which depends on a fixed internal degree of freedom $s_a$, which may take on one of two values to allow for a system to be put into a superposition state. In the massive case, this internal degree of freedom may correspond to the spin of a particle whereas for light, we shall take it to correspond to its polarisation state. Here $\ket{+/-}$ denotes left/right-handed circular polarisation. The internal configuration between two of the four different branches is then described by a state $\ket{\sigma}=\bigotimes_a \ket{s_a}$. After passing the light pulses through the beam splitter separating light into horizontal or vertical polarisation, a quarter-waveplate may be used to correspondingly have right-handed and left-handed circular components in output modes. At the end of the gravitational phase induction, the light pulses pass again through quarter-waveplates before passing through a beam splitter to detect the entanglement.
Per definition, each beam splitter is assumed to transmit and reflect equally with no losses in our setup. 

Eventually, after having successfully put each of the two light pulses in a spatial superposition correlated with their polarisation degree of freedom, each branch will pick up a gravitationally generated phase. Even though light beams in coherent states~\cite{Rosas_Ortiz_2019,Gerrits_2010,Glancy_2008,Sanders_2012} may contain a high number of photons which allows us to source an appreciable metric perturbation, let us state right now that they cannot be used for the GME protocol. Coherent states (with no squeezing) incident on a beam splitter do not become entangled ~\cite{QOKnight}. As such, in the remainder of this work, we focus on states for which the beam splitter is entangling, and we will discuss such states in a later section. At the end of the gravitational interaction stage, the joint spatial-polarisation subsystem state will have evolved to
\begin{equation}
\begin{aligned}
\ket{\psi(\tau)}&=\frac{1}{2}\left( e^{i\phi_{\text{L}\text{L}}(\tau)}\ket{\text{L};+}_1\ket{\text{L};-}_2+e^{i\phi_{\text{L}\text{R}}(\tau)}\ket{\text{L};+}_1\ket{\text{R};+}_2
+e^{i\phi_{\text{R}\text{L}}(\tau)}\ket{\text{R};-}_1\ket{\text{L};-}_2+ e^{i\phi_{\text{R}\text{R}}(\tau)}\ket{\text{R};-}_1\ket{\text{R};+}_2\right)\\
&\propto\frac{1}{2}\left(\ket{\text{L}\text{L};+-}+e^{i\Delta\phi_{\text{L}\text{R}}}\ket{\text{L}\text{R};++}
+e^{i\Delta\phi_{\text{R}\text{L}}}\ket{\text{R}\text{L};--}+ \ket{\text{R}\text{R};-+}\right),\\
\end{aligned}
\label{eq:entangledstate}
\end{equation}
where we have left unspecified a global phase, we have defined $\Delta\phi_{\text{L}\text{R}}:=\phi_{\text{L}\text{R}}-\phi_{\text{L}\text{L}}, \Delta\phi_{\text{R}\text{L}}:=\phi_{\text{R}\text{L}}-\phi_{\text{L}\text{L}}$ and used $\phi_{\text{R}\text{R}}=\phi_{\text{L}\text{L}}$, and $\tau$ is the travel time in the longitudinal direction.
The gravitational interaction prior to the first light pulse impinging on the mirror is disregarded, as the splitting and refocusing durations in the transverse direction are considered negligible compared to the travel time in the longitudinal direction, where the interaction takes place. This assumption is similar to assuming a negligible duration for splitting and refocusing in the massive GME protocol case. Furthermore, through the action of appropriate waveplates, one can always ensure that the four different branches are all described by either left or right-handed circular polarisation. For example, circularly polarised light can be produced by passing linearly polarised light through a quarter-waveplate, and vice versa~\cite{SalehPhotonics}. This circular polarisation assumption simplifies the expression of the stress-energy density, which otherwise in the longitudinal polarisation case, is modulated along the propagation direction.

\subsection*{Path integral description}
The original GME protocol assumed Newtonian gravitational action-at-a-distance, but to make the locality of the gravitational field present, a manifestly Lorentz covariant approach using path integrals~\cite{Rosenfelder2012} known from quantum field theory was first invoked by Christodoulou et al~\cite{Brukner_2022}. We will adopt this treatment due to its freedom of choice in inserting the desired ingredients coupling to each other gravitationally via the action, further details are given in their paper. Given that the internal degree of freedom is entangled with the output spatial mode, the time evolution operators is expressed as $U_{i\rightarrow f}=\sum_{\sigma}\ket{\sigma}\bra{\sigma}\otimes U^{\sigma}_{i\rightarrow f}$
where $U^{\sigma}_{i\rightarrow f}$ is defined from the initial and final total states $\ket{\Psi^{i,f}}$ of the system and fields. Here it is assumed that the field and the systems are not entangled with each other at the beginning and end of the protocol. The assumptions from Ref.~\cite{Brukner_2022} are invoked, that is $(1)$ gravity is weak, $(2)$ the beam splitter does not couple to the gravitational field and the beam splitter action is only correlated with the spatial motion of the polarisation state, $(3)$ stationary phase approximation (no loop corrections) and $(4)$ second stationary phase approximation and orthogonality of spatial states of the different branches. Then the unitary time-evolution operator corresponding to a polarisation configuration $\sigma$ evaluates to $U^{\sigma}_{i\rightarrow f}\propto \exp{\left(\frac{iS^{\sigma}}{\hbar}\right)}\ket{\Psi^f}\bra{\Psi^i}$, where $S^{\sigma}$ is the joint on-shell action. There exists a decomposition of the action into $S=S_0+S_{\mathcal{F}}$ with $S_0$ serving as a global phase coincident for all internal degree of freedom configurations $\sigma$ and a phase which depends on the field mediation $S_{\mathcal{F}}$. Finally, the phase is expressed as~\cite{Brukner_2022}:
\begin{equation}
    \phi_{\sigma}=\frac{S^{\sigma}_{\mathcal{F}}[x^{s_a}_a, \mathcal{F}[x^{s_a}_a]]}{\hbar}.
\end{equation}

\subsection*{Action for linearised gravity}
\label{section:Action for linearised gravity}
The gravitational field sourced by systems suitable for the GME protocol can be described by a weak perturbation of the Minkowski background metric $\eta_{\mu\nu}$. In the following, the action for such a linearised theory of gravity may be derived using standard general relativity~\cite{Carroll_2019}. Weak gravitational fields $\mathcal{F}$, such as those sourced by light, are described by a metric
$g_{\mu\nu}=\eta_{\mu\nu}+h_{\mu\nu}$ with $\abs{h_{\mu\nu}}\ll 1$ and signature $(-,+,+,+)$. The on-shell action, after integration by parts, then takes the form~\cite{Brukner_2022}:
\begin{equation}
\label{eqn:action_onshell}
    S^{\sigma}_{\mathcal{F}}=\frac{1}{4}\int d^4 x h_{\mu\nu}T^{\mu\nu},
\end{equation}
where $T_{\mu\nu}$ is the energy-momentum tensor. Here we work in the Lorenz gauge $\partial^{\nu}\bar{h}_{\mu\nu}=0$, where we denote the trace reversal of a tensor $h_{\mu\nu}$ by $\bar{h}_{\mu\nu}=h_{\mu\nu}-\frac{1}{2}\eta_{\mu\nu}h$ and $h=\eta_{\mu\nu}h^{\mu\nu}$ as the trace of $h_{\mu\nu}$. The retarded solution~\cite{Raetzel_2016, Brukner_2022} of the linearised Einstein field equations is
\begin{equation}
\begin{aligned}
    h_{\mu\nu}(t,\vec{x})=\frac{4G}{c^4}\int dx'^3 \eta_{\mu\alpha}\eta_{\nu\beta}\frac{\bar{T}^{\alpha\beta}(\vec{x}',t_r)}{|\vec{x}-\vec{x}'|},
\label{eq:retarded_sol}
\end{aligned}
\end{equation}
where the retarded time $t_r=t_r(t,\vec{x},\vec{x}')$ is defined by $t_r:=t-|\vec{x}-\vec{x}'|/c$.

\subsection*{Stress-energy tensor and metric perturbation for a single light pulse}
Before we can determine the action in the photonic GME protocol, in this section we give a brief overview of useful calculations which were undertaken in previous work by Rätzel et al~\cite{Raetzel_2016} investigating the effect of the gravitational field sourced by a monochromatic laser pulse of length $L$ on another test pulse, in which diffraction and overlap are disregarded.

Consider, in free space and flat spacetime~\cite{WheelerGravitation_1973} an electromagnetic plane wave propagating in vacuum with energy density $U$. We assume it is propagating in the $\hat{z}$-direction, thus the only non-vanishing components of its associated stress-energy tensor are $T^{00}_0=T^{0z}_0=T^{z0}_0=T^{zz}_0=U$. The stress-energy tensor takes the form ${T_0}^{\mu\nu}=A u(ct-z)\delta(x)\delta(y)$, where $A$ is the effective area of the pulse in the $xy$-plane and $u(ct-z)$ is the energy density of the pulse along the $z$-axis, see Ref.~\cite{Raetzel_2016} for details.

Consequently, the $00$-component of Eq.~\eqref{eq:retarded_sol} reads
$h^p_0=\frac{4GA}{c^4}\int dz' \frac{u(ct_r(x,y,z,t,z')-z')}{\sqrt{\rho(x,y)^2+(z'-z)^2}}$, where the retarded time is $t_r=t-\sqrt{\rho(x,y)^2+(z-z')^2}/c$ and $\rho(x,y)=\sqrt{x^2+y^2}$.
Via the variable substitution $\zeta(x,y,z,z')=(z'-z)+\sqrt{\rho(x,y)^2+(z'-z)^2}$ Rätzel et al. arrive at
\begin{equation*}
h^p_0= \frac{4GA}{c^4}\int_{\zeta(x,y,z,a)}^{\zeta(x,y,z,b)} d\zeta \frac{u(ct-\zeta-z)}{\zeta}.
\end{equation*}

The integration boundaries $a, b$ of each pulse contribution arise from the intersection of the world sheet boundaries of each pulse with the past light cone $J^{-}$ of an observer located at spacetime position $x^{\mu}=(x,y,z,t)$~\cite{Raetzel_2016}. For pulses of length $L$, the auxiliary integration boundaries $\bar{a},\bar{b}$ are defined as solutions of $t_{r}(z')=\frac{z'+L}{c}$ and $t_{r}(z')=\frac{z'}{c}$, which are given by
\begin{equation}
    \bar{a}(x,y,z,t)=z+\frac{(ct-L-z)^2-\rho(x,y)^2}{2(ct-L-z)} \  \text{ and } \  \bar{b}(x,y,z,t)=z+\frac{(ct-z)^2-\rho(x,y)^2}{2(ct-z)}.
\end{equation}

The actual integration boundaries $a,b$ are then defined by
\begin{equation*}
  [a,b]=
    \begin{cases}
      \emptyset, &\bar{a}<\bar{b}<0<D\qquad \text{(Zone $I_{-}$)}\\
      \emptyset, &0<D<\bar{a}<\bar{b}\qquad\text{(Zone $I_{+}$)}\\
      [0,\bar{b}], &\bar{a}<0<\bar{b}<D\qquad \text{(Zone $II$)}\\
      [\bar{a},\bar{b}], &0<\bar{a}<\bar{b}<D\qquad \text{(Zone $III$)}\\
      [\bar{a},D], &0<\bar{a}<D<\bar{b} \qquad \text{(Zone $IV$)}\\
      [0,D], &\bar{a}<0<D<\bar{b}\qquad \text{(Zone $V$)}\\
    \end{cases}       
\label{eq:integration_bound}
\end{equation*}

\begin{figure}
    \centering
    \includegraphics[scale=0.75]{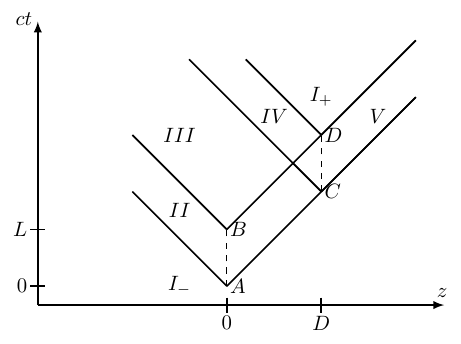}
    \caption{Sketch of the spacetime diagram of a single pulse in the $tz$-plane. The various zones define different metric perturbations and arise from the intersection of the pulse's world sheet with the past light cone $J^{-}$ of a spacetime event $x^{\mu}$. The world sheet of the single pulse is spanned by $A,B,C,D$. Figure adapted from Ref.~\cite{Raetzel_2016}.}
    \label{fig:diagram_onepulse}
\end{figure}

The different zones are shown in Fig.~\ref{fig:diagram_onepulse}. Zones $I_{\pm}$ are causally disconnected from the pulse and thus the metric is not perturbed. Zone $II$ is defined by the pulse emission from the mirror, zone $III$ only describes the passage of the pulse (excluding emission and absorption), zone $IV$ describes pulse absorption only, while zone $V$ describes both emission and absorption. The world sheet of the pulse is spanned by points $A,B,C$ and $D$ in Fig.~\ref{fig:diagram_onepulse}.\\

For circularly polarised pulses, inserting the constant energy density $u(ct-z)=u_0$ into the expression for $h^p_0$ yields
\begin{equation}
\begin{aligned}
    h^p_0[x,y,z,t]=
    \frac{4GA}{c^4}u_0\ln{\left(\frac{\zeta(x,y,z,b_{x,y,z,t})}{\zeta(x,y,z,a_{x,y,z,t})}\right)}, 
\end{aligned}  
\label{eq:hp_expr}
\end{equation}
where we have used an index notation for legibility $a_{x,y,z,t}:=a(x,y,z,t)$ and similarly for $b$.

\subsection*{Stress-energy tensor, metric perturbation and action for two pulses}
\begin{figure}
    \centering
    \includegraphics[scale=1]{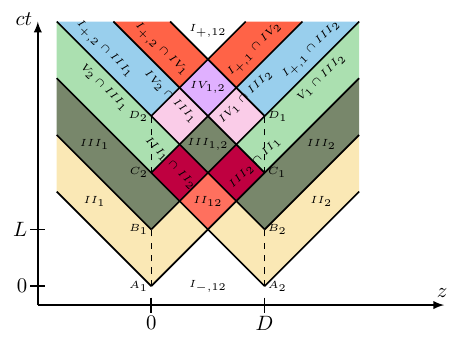}
    \caption{Sketch of the spacetime diagram of two counter-propagating pulses projected onto the $tz$-plane. The two pulses have equal $y$ coordinates, but different $x$ coordinates. The various zones are found in analogy to the single pulse case.}
    \label{fig:diagram_twopulse_bang}
\end{figure}

Having presented the useful notation and derivation of Rätzel et al. for the stress-energy tensor and metric perturbation for a single light pulse, we will now slightly extend the above considerations to cater to the discussion for the case of two counter-propagating laser pulses being present, each with stress-energy tensor $T_a^{\mu\nu}$, where $a\in\{1,2\}$ is a pulse label. If the effective size of the two pulses is much smaller than their separation, diffraction effects are insignificant and the total stress-energy tensor takes the form
\begin{equation}\label{eq:Tfor2pulses}
    T^{\mu\nu}(t,\vec{x})=\sum_{a=1}^{2}T_a^{\mu\nu}=A u_0\sum_{a=1}^{2}\delta(x-x_a)\delta(y-y_a),
\end{equation}
where $A$ is the effective area of each pulse in the transverse plane and $(x_a,y_a)$ are the respective positions of the pulses in the $xy$-plane. For two counter-propagating light pulses propagating along the $\hat{z}$-direction, initially separated by a distance $D$ with a sufficient separation along the $x$-axis with $x_2-x_1\neq 0$, we set $y_1=y_2=const$. The spacetime diagram for two counter-propagating pulses projected onto the $tz$-plane is plotted in Fig.~\ref{fig:diagram_twopulse_bang}.

In the context of linearised gravity, metric perturbations due to each pulse may be simply added. In this way, interference effects have been implicitly ignored.

\begin{figure}
    \includegraphics[scale=0.6]{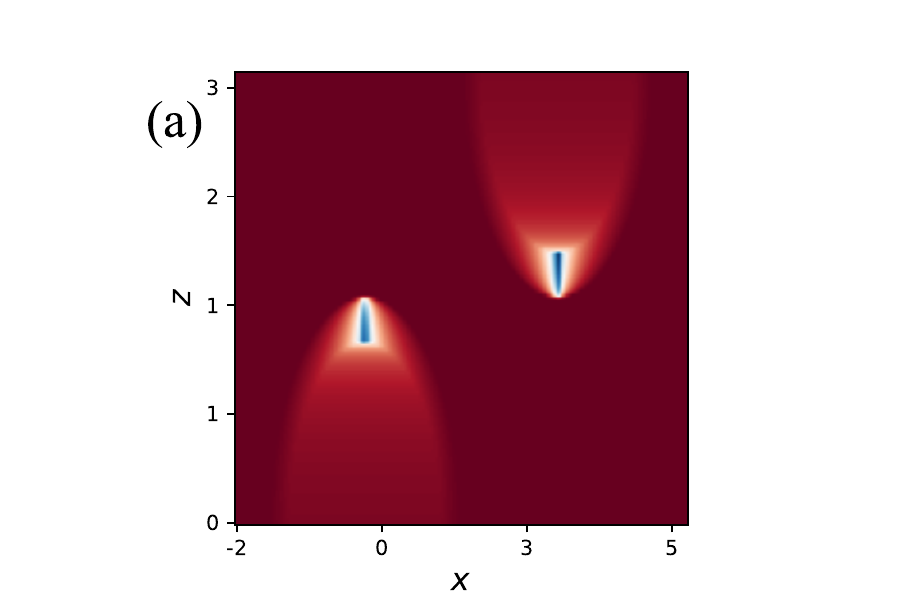}
    \includegraphics[scale=0.6]{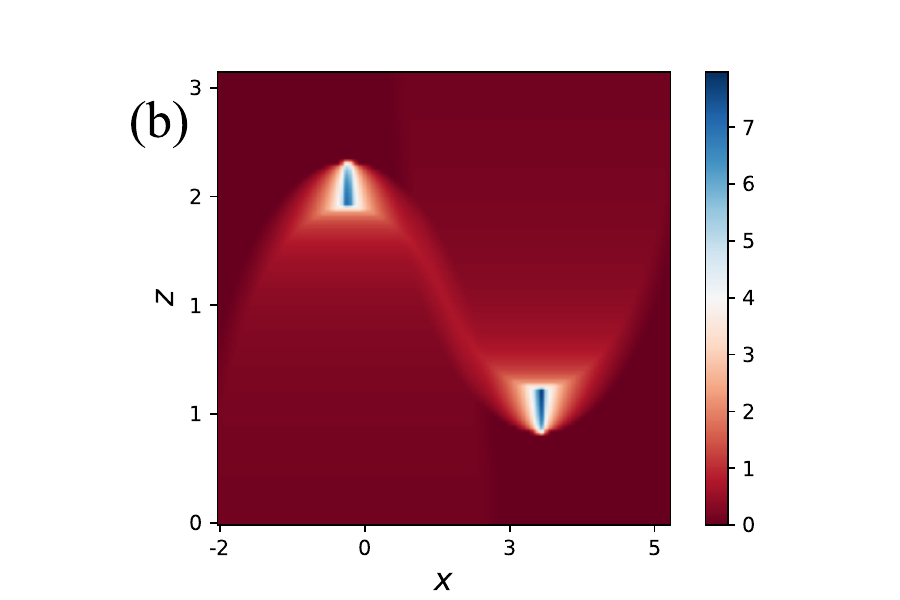}
    \caption{The total metric perturbation $h^p[x,0,z,t]$ for two counter-propagating light pulses located at $x_1=\Delta x,x_2=d$ in the $xz$-plane is shown for (a) $ t= D/(2c)$, (b) $t=D/c$. We have set $D=3\text{m}, L=0.3\text{m}$ and $\frac{4GA}{c^4}u_0= c= 1$. Lengths are in units of $ct$.}
    \label{fig:hptotal_exampleplot}
\end{figure}

For two counter-propagating pulses along the $\hat{z}$-direction that are each located at $x_{1,2}$ along the $x$-axis, we obtain the total contribution to the metric by substituting the single pulse metric contributions from Eq.~\eqref{eq:hp_expr}:
\begin{equation}
\begin{aligned}
    h^p[x,y,z,t]&= h_1^p[x,y-y_1,z,t]+h_2^p[x,y-y_2,z,t]=h^p_0[x-x_1,y-y_1,z,t]+h^p_0[x-x_2,y-y_2,-z+D,t].
\end{aligned}
\end{equation}
The total metric perturbation $h^p$ for two counter-propagating light pulses in the $xz$-plane is shown in Fig.~\ref{fig:hptotal_exampleplot}.

We are now in a position to express the action associated with the process at hand. By substituting the expression for $T^{\mu\nu}$ into Eq.~\eqref{eqn:action_onshell}, one can establish

\begin{multline}
S^{\sigma}_{\mathcal{F}}=\kappa\int_{z=-\infty}^{\infty}dz\int_{t=0}^{\infty}dt \ln{\left(\frac{\zeta(0,0,z,b_{0,0,z,t})}{\zeta(0,0,z,a_{0,0,z,t})}\right)}+
\ln{\left(\frac{\zeta(x_1-x_2,0,-z+D,b_{x_1-x_2,0,-z+D,t})}{\zeta(x_1-x_2,0,-z+D,a_{x_1-x_2,0,-z+D,t})}\right)}\\ + 
\ln{\left(\frac{\zeta(x_2-x_1,0,z,b_{x_2-x_1,0,-z,t})}{\zeta(x_2-x_1,0,z,a_{x_2-x_1,0,z,t})}\right)}+
\ln{\left(\frac{\zeta(0,0,-z+D,b_{0,0,-z+D,t})}{\zeta(0,0,-z+D,a_{0,0,-z+D,t})}\right)},
\label{eq:action}
\end{multline}
where $\kappa:=4GA^2 u_0^2/c^4$, see Methods for details. The values for $x_{1,2}$ used for the different spatial branches are stated in Table~\ref{table:config}.

\begin{table}[h!]
\centering
\begin{tabular}{ |p{3cm}||p{3cm}|p{3cm}|p{3cm}|  }
\hline
 ~& $x_1$&$x_2$\\
 \hline
 \hline
 $\text{L}\text{L}$ & $0$ & $d$\\
 \hline
 $\text{L}\text{R}$ & $0$ & $d+\Delta x$\\
 \hline
 $\text{R}\text{L}$ & $\Delta x$ & $d$\\
 \hline
 $\text{R}\text{R}$ & $\Delta x$ & $d+\Delta x$\\
 \hline
\end{tabular}
\caption{The values for $x_{1,2}$ in Eq.~\eqref{eq:action} are shown for the different spatial branches $\text{L}\text{L}, \text{L}\text{R}, \text{R}\text{L}, \text{R}\text{R}$.}
\label{table:config}
\end{table}

\subsection*{Numerical estimation of the phase evolution}
From the expression of the action, we can now proceed to find the phase evolution associated with the counter-propagation of light pulses in each pair of modes. Given that $h^p_0[x,\cdot]=h^p_0[-x,\cdot]$, we consider the relative phase factor $e^{i\Delta\phi_{\text{R}\text{L}}}$ for the pair of modes that are the closest from one another $\ket{\text{R}\text{L}}$:
\begin{equation}\label{eq:phasephiRL}
\begin{aligned}
    \hbar\Delta\phi_{\text{R}\text{L}}&=A u_0\int_{z=-\infty}^{\infty}dz\int_{t=0}^{\infty}dt\left(h^p_0[d-\Delta x,0,-z+D,t]+h^p_0[d-\Delta x,0,z,t]-h^p_0[d,0,-z+D,t]-h^p_0[d,0,z,t]\right).
\end{aligned}
\end{equation}

\begin{figure}[h]
    \centering
    \includegraphics[scale=0.6]{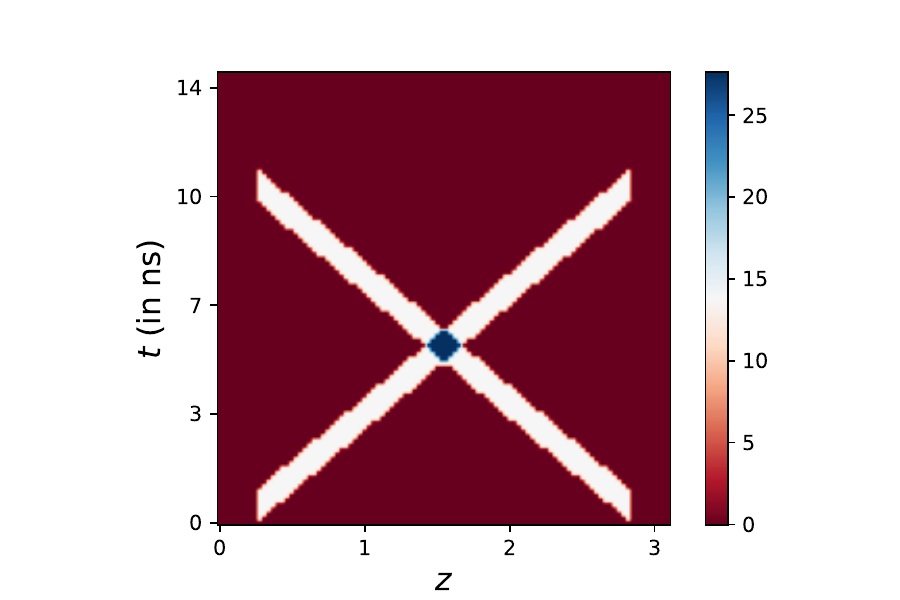}
    \caption{Plot of the integrand of the relative phase at closest approach $\Delta\phi_{\text{R}\text{L}}$ for two counter-propagating pulses with $D=3\text{m}$ and $L=0.3\text{m}$ in the $tz$-plane. We have set $\frac{4GA}{c^4}u_0= c =1$.  Lengths are in units of $ct$.}
    \label{fig:totalhp_integrand}
\end{figure}
To evaluate Eq.~\eqref{eq:phasephiRL}, one has to numerically integrate $h^p_0[x,y,z,t]$ with sufficient accuracy, where the $z$-integration limits will effectively be finite and increasing with an \textit{effective} interaction time $\tau_{\text{eff}}$, which will serve as an upper bound to the time integral, as we shall now argue. As we can see in Fig.~\ref{fig:totalhp_integrand} the integrand of Eq.~\eqref{eq:phasephiRL}, vanishes for values of $z$ outside of the lightcone of the spacetime events of emission of the two light pulses. Entanglement is not generated for events that are spacelike separated, in accordance with previous findings~\cite{Brukner_2022}. Furthermore, the integral of the relative phase converges as the effective interaction time $\tau_{\text{eff}}$ is increased. This is, however, only a finite size effect due to the length of the interferometer $D$ limiting the gravitational interaction. 

The integration itself is done by discretising the integral into a coarse-grained sum. As the complete, numerically accurate phase evolution calculation requires incorporating large sections of the full spacetime integral $\int_{z=-\infty}^{\infty}\int_{t=0}^{\infty} dt dz$ for which no substantial phase is accumulated, we choose to restrict the integration domain to $0\leq t\leq2D/c$ and $-D-L\leq z\leq D+L$, a spacetime region encapsulating the pulse crossing and mutual interaction process, which will give a reasonable approximation to the actual accumulated phase. The effect of metric perturbations outside of this spacetime region was shown to be negligible for different sets of parameters and can thus be safely ignored. To be explicit, we have chosen $\tau_{\text{eff}} = 2D/c$ which accordingly allows to restrict the integral over $z$ so that Eq.~\eqref{eq:phasephiRL} may be approximated by

\begin{equation}\label{eq:approxphase}
\begin{aligned}
    \hbar\Delta\phi_{\text{R}\text{L}}&\approx A u_0\int_{z=-c\tau_{\text{eff}}/2-L}^{c\tau_{\text{eff}}/2+L}dz\int_{t=0}^{\tau_{\text{eff}}=2D/c}dt\left(h^p_0[d-\Delta x,0,-z+D,t]+h^p_0[d-\Delta x,0,z,t]-h^p_0[d,0,-z+D,t]-h^p_0[d,0,z,t]\right).
\end{aligned}
\end{equation}

In the massive GME protocol, the particles are kept at constant separation, hence a longer interferometer length $D$ gives a larger relative phase accumulation. In contrast, increasing $D$ in the photonic protocol, where the light pulses are counter-propagating, does not give the same effect as most of the relative phase accumulation occurs when the pulses are close to each other. 

Let us now give an example of a set of parameters allowing for an appreciable relative phase. For the calculation to be valid, we remind that diffraction and interference effects must be negligible, that is, we must operate in the regime $d-\Delta x\gg\lambda$ where $\lambda$ is the wavelength. We will conservatively set $\lambda$ such that $d-\Delta x=10\lambda$. To give an order of magnitude estimate on the feasibility of the protocol, we will enforce $\Delta\phi_{RL}\sim 1 \text{rad}$ and determine corresponding parameters for which this can be achieved. In the discussion, we will comment on the current feasibility and the challenges to overcome.

Considering typical and widely available nanosecond pulsed lasers, we assume a central wavelength $\lambda= 1\,\mu \text{m}$ and pulse length $L=0.3\,\text{m}$. Choosing a splitting distance $\Delta x= 1\,\text{cm}$ and satisfying $\Delta x=d-10\lambda$ in order to have negligible beam overlap, we also choose a typical laboratory friendly interferometer length $D$ on the order of a few meters so that $D/L\approx 10$. Since for a laser with power $P$ we have $P=A u_0 c$, we can write $\kappa=4GA^2 u_0^2/c^4=4GP^2/c^6$. According to numerics, in order to achieve an appreciable relative phase of $\Delta\phi_{RL}\approx 0.29\, \text{rad}$, which would correspond to a state with negativity $\mathcal{N} \approx 0.072$ we require a power $P\approx 30\, \text{PW}$. Such a pulse contains on average $\ev{N}=PL\lambda/(hc^2)\approx 10^{26}$ photons, an admittedly demanding number, as we shall discuss in the next sections. This estimate may be refined by considering higher-order contributions in the quantum field theoretic formulation, and accounting for the bandwidth of the pulse. Although the power requirement to obtain a phase of order $1$ is extreme, if we are able to reliably detect entanglement with a much lower phase signal, then the required photon number may be decreased. We undertake such an analysis in the following section.

\subsection*{{Entanglement witnessing in practice}}
Investigating the detection of entanglement with lower phase signal is important for the presented protocol, as this allows to consider less extreme photon numbers, making the experiment more realistic, and bringing the regime further away from the Schwinger electron-positron generation limit~\cite{zhu2016dense} of $10^{29} \text{W}\cdot\text{cm}^{-2}$. Our first estimation corresponds to a power density of $10^{24} \text{W}\text{cm}^{-2}$ assuming a waist cross section of $1\mu{\text{m}}^2$. Detecting entanglement is only possible by building reliable statistics, and detecting a lower signal asks for good repeatability and low noise, which are the main strengths of the photonic protocol.

An efficient way of certifying the bipartite entanglement is by measuring an entanglement witness $W$, an operator that satisfies $\Tr[W\rho]\geq 0$ for any separable state $\rho$. An optimal entanglement witness reads~\cite{Chevalier_2020} $W=\mathbbm{1}\otimes\mathbbm{1}-X\otimes X - Z\otimes Y - Y\otimes Z$, where $\{\mathbbm{1},X,Y,Z\}$ forms the set of single qubit Pauli operators. Note that the spectrum of $W$ is $\{-2,2\}$. In the rest of this work, we shall use standard computational basis notations for the field modes $\ket{L,R}$ and their associated polarisation state, that is we write $\ket{L,R}$ as $\ket{0,1}$. In the noiseless scenario, the witness expectation value is simply $\Tr[W\ket{\theta}\bra{\theta}]=1-\left(\sin{(\Delta\phi_{\text{L}\text{R}})}+\sin{(\Delta\phi_{\text{R}\text{L}})}\right)-1/2\left(1+\cos{(\Delta\phi_{\text{L}\text{R}}-\Delta\phi_{\text{R}\text{L}})}\right)$, where $\ket{\theta}$ is the bipartite polarisation state. This witness will pick up entanglement, in the noiseless case, as soon as the relative phase accumulation is appreciable.

In practice, the entanglement cannot be certified in a single-shot experiment. One instead has to repeat the witness measurement and carry out hypothesis testing. The witness measurement has corresponding POVM elements $\{\dyad{\Pi}, \mathbbm{1}-\dyad{\Pi}\}$ where $\ket{\Pi} = \frac{1}{2}(\ket{00} + i\ket{01} + i\ket{10} + \ket{11})$ is the eigenstate associated to the negative eigenvalue of $W$. As indicated in Ref.~\cite{blume2010entanglement}, the empirical data $\mathcal{D}$ on which the hypothesis testing is carried out may be represented by the frequency ${f} = n_{\text{yes}}/N$ where $n_{\text{yes}}$ is the number of times the negative eigenvalue outcome has occurred (i.e. the witness has detected entanglement), among a total of $N$ repetitions. The likelihood $\mathcal{L}(\rho)$ of a state $\rho$ given the observation of data $\mathcal{D}$ is the conditional probability $\mathbb{P}(\mathcal{D}|\rho)$ of obtaining such data from measurements on state $\rho$. In order to certify that the state at hand is entangled, one may build the likelihood ratio $\Lambda = \underset{\rho\in\mathcal{D}_{\text{sep}}}{\text{max}}\mathcal{L}(\rho)/\underset{\rho\in\mathcal{B}(\mathcal{H})}{\text{max}}\mathcal{L}(\rho)$, where $\mathcal{D}_{\text{sep}}$ is the set of separable states and $\mathcal{B}(\mathcal{H})$ is the complete state space. High values of the logarithmic likelihood ratio $\lambda = -2\log_2(\Lambda)$ are in support of entanglement. It is known~\cite{blume2010entanglement} that in the case of entanglement witnessing $\lambda = 2N(1 - H(f))$ where $H(f) = -f\log_2(f) - (1-f)\log_2(1-f)$ is the data's entropy, and furthermore that $\lambda$ will behave like a semi-$\chi_1^2$ random variable for a large enough number of repetitions (i.e. when the separable space behaves like a half space with respect to the spread of the maximum likelihood estimators). Explicitly, we aim to seek what value of $\lambda$ is ``large enough'', and what value of $N$ is large enough in order to consistently obtain a large enough $\lambda$. To determine a critical value for $\lambda$, one must define a desired significance level $\alpha$, or equivalently a confidence level $(1-\alpha)$. For the purpose of our investigation, we shall work with the confidence level $1-\alpha = 0.95$. This amounts to fixing a critical log-likelihood ratio value $\lambda_c$ such that carrying out the likelihood ratio test given some data $\mathcal{D}$ generated by the most likely separable state will yield a value $\lambda > \lambda_c$ with probability $0.05$. In other words, we require a false positive rate of $5\%$, which accordingly fixes $\lambda_c$. Numerically, we simulate $10^5$ witness measurements on the initial separable state to obtain a given $\lambda$, and repeat the process $10^4$ times. Then $\lambda_c$ is chosen to be the $95$-th percentile of the obtained values of $\lambda$ on the separable state. Once $\lambda_c$ is determined, one can carry out witness measurements on the state of interest. Whenever a value of $\lambda > \lambda_c$ is obtained, the experiment has successfully certified entanglement with a confidence level of $95\%$. Let us now consider some figure of merit on the number of repetitions of the experiment, in order to have appreciable success rates. In particular, we expect that with lower phase signals (due to power limitations for instance), the success rates with a few repetitions will rapidly decay, as the state will become closer to the separable boundary. The required number of witness measurements to maintain a good success rate as the phase signal decreases is of central interest.

\begin{figure}
    \centering
    \includegraphics[scale=0.6]{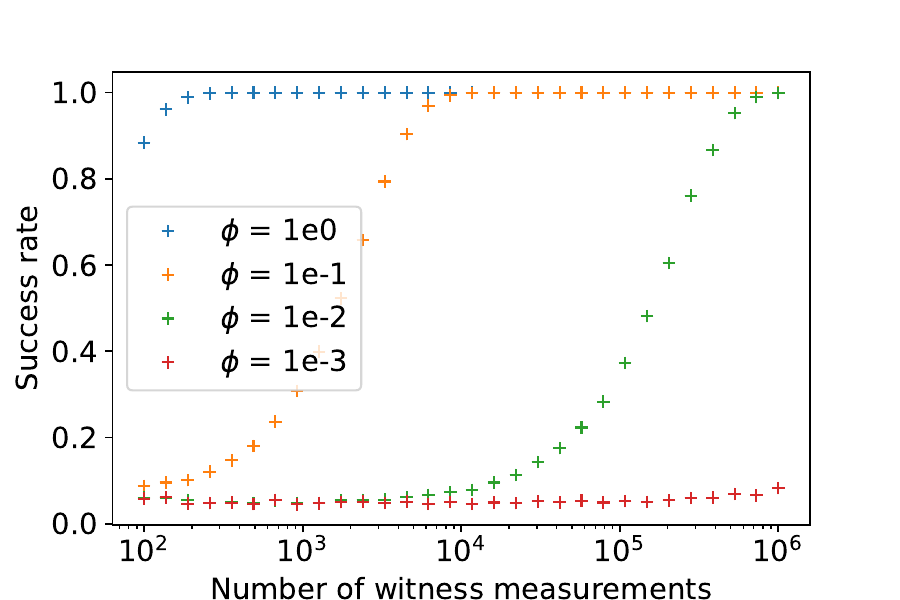}
    \caption{Success rates with respect to the number of repetitions of the witness measurement, for different orders of magnitude of phase signal $\phi\in\{1, 10^{-1}, 10^{-2}, 10^{-3}\}$, where the approximate state negativities are $\mathcal{N}\in2.5\times \{10^{-1}, 10^{-2},  10^{-3}, 10^{-4}\}$.}
    \label{fig:SuccessRates}
\end{figure}

In order to simplify our analysis, we approximate the noiseless bipartite polarisation state by $\ket{\theta(\phi)} = \frac{1}{2}(\ket{00} + \ket{01} + e^{i\phi}\ket{10} + \ket{11})$, where $\phi = \Delta\phi_{RL}$. This amounts to neglecting all but the fastest accumulating phase. We shall denote the corresponding density operator $\rho(\phi) = \dyad{\theta(\phi)}$. The results shown in Fig.~\ref{fig:SuccessRates} confirm that the required number of measurements increase as $\phi$ decreases. We observe that for $\phi = 1 \text{rad}$ a few $10^2$ repetitions is enough to consistently yield a successful outcome. For $\phi = 0.1 \text{rad}$ roughly $10^4$ repetitions are required, and $10^6$ repetitions are needed when $\phi = 0.01 \text{rad}$. For $\phi = 1 \text{mrad}$ a million witness measurements will rarely ever succeed in certifying entanglement at the $95\%$ confidence level. Numerics reveal that for $10^8$ witness measurements, the success rate is above $99\%$. In fact, decreasing the phase signal by an order of magnitude while maintaining a similar success rate requires to increase the number of measurements by two orders of magnitude. Indeed, measuring the witness is equivalent to measuring the Boolean observable $B = \frac{1}{2}(\mathbbm{1} - \frac{W}{2})$ with eigenvalues $\{0,1\}$, where the value $1$ is associated with the negative measurement outcome of $W$ (supporting entanglement). Our witness empirical data is essentially $f = \frac{1}{N}\sum_{i=1}^N B_i$, where $B_i$ are independent measurement outcomes, which by the central limit theorem converges in distribution to a Gaussian centered around $\Tr(B\rho(\phi))$ with characteristic width $\frac{1}{\sqrt{N}}$. By considering small values of $\phi$, for which one may write the first order approximation $\Tr(W\rho(\phi)) = -\phi + O(\phi^2)$, one arrives at $\Tr(B\rho(\phi)) = \frac{1}{2}+\frac{\phi}{4} + O(\phi^2)$. Note that if the lowest order term was superlinear in $\phi$ the required number of repetitions $N$ would grow much faster when the phase signal is reduced.

In light of these observations, one can now roughly estimate a lower required photon number for the experiment to be successful. Let us suppose that the interferometer length is $D\sim 1 \text{m}$, which corresponds to a travel time $\tau \sim 10 \text{ns}$. Then the entanglement witness measurement rate may be as high as $10^8$ measurements per second, or $10^{13}$ per day. This could allow for consistent successful experiments spanning ten days with a dominant phase signal as low as $10^{-6}\text{rad}$, bringing down the photon number requirement by $3$ orders of magnitude, that is $\ev{N} \sim 10^{23}$ photons, or a power $P\approx 30\text{TW}$.

\section*{Discussion}
Recent proposals aiming to witness gravity mediated entanglement between two massive objects~\cite{Bose_2017,Marletto_2017} have given a second wind to the age-old investigation of unified quantum gravity. While substantial work around the original protocols has refined and improved certain aspects, in this work we have chosen to analyse a photonic variant despite the well-known weak coupling of light to gravity. We considered the basics of the experimental implementation using two Mach-Zehnder interferometers and derived the metric perturbation sourced by two counter-propagating light pulses using linearised gravity. Through the path integral formalism, we have derived the action of the pulses leading to gravitational phase induction, and for some parameters we have given a figure of merit of the average photon number required for a order unity phase signal. By harnessing the natural repeatability of photonic experiments, in combination with an optimal entanglement witness, we have illustrated how one may lower the required phase signal.

The advantages of considering a photonic version of the GME protocol are manifold. First and foremost the usage of light instead of matter relieves us of unwanted and otherwise unavoidable entangling effects such as dipole-dipole interactions or Casimir coupling, and difficulties in controlling the motion of massive objects such as microdiamonds with an NV center. Second, this doubly interferometric setup may also serve to demonstrate gravitational coupling between light beams, through dephasing, rather than spatial deflection. Thirdly, a photonic experiment allows us to build up statistics more reliably than with matter, and the certification of entanglement unavoidably will require many repetitions of bipartite Pauli measurements. Finally, the photonic GME protocol is a fully relativistic test of GME, and as such may be seen as a more convincing test for the quantum nature of gravity. The path integral formalism used in the analysis of the entanglement generation also allows for a manifestly local description of gravity thus strengthening the argument for certifying its non-classicality further.

This being said, the implementation of the photonic GME protocol is by no means straightforward. To formulate more precise claims on its feasibility, we lay out some further questions that may constitute interesting future investigations.

The most critical drawback of using light in the GME protocol is of course the weakness of the gravitational effects of light, and so the need for quite intense light pulses. It would be tempting to think that current laser technology allowing for petawatt ultrashort pulses~\cite{liu2020house} may already allow the implementation of the protocol. However, one is then faced with at least two issues. First is the control of high-intensity light pulses, which are known to damage the optical components constituting the whole apparatus. Second, and much more problematic, is the fact that standard high-power laser pulses are not usable for they are classical light states (in the sense that their Glauber-Sudarshan P function is positive), and as such are not entangled through beam splitting. Instead one has to resort to non-classical light, for which high average photon number states are much harder to achieve. Ideally for the GME protocol, one would like to achieve the maximally entangled state after beam splitting in each interferometer $\ket{\psi}_{\text{out}}=\frac{1}{\sqrt{2}}\left(\ket{N}_x\ket{0}_z+\ket{0}_x\ket{N}_z\right)$, which is commonly called the \textit{NOON} state. Due to the high interest for use in quantum information science, several protocols have been proposed to generate NOON states~\cite{Kok2002,Hofmann2004,MsCusker2009,Gerry_2010,Eaton2021,Guha_Majumdar_2022}. Unfortunately, current high-NOON states contain $N<10$ photons, and as such the generation of an appreciable metric perturbation from a NOON state is still a distant dream.

Nevertheless, in our view, the photonic protocol should not be discarded without some further analysis. As we have shown, having a very low single-shot phase signal does not necessarily constitute a dead-end. The absence of non-gravitational entangling forces between the light pulses, the use of an optimal entanglement witness, and the rate at which measurements can be repeated, all help alleviate the minuteness of the signal. While current NOON states are very weak, this limitation is not fundamental, barring any spontaneous collapse process a la Penrose~\cite{penrose1996gravity} which would in any case also affect the matter GME protocol. Indeed, the approach presented in Ref.~\cite{Afek2010} to generate NOON states with large $N$ is, in principle, scalable up to arbitrarily high photon numbers. Approximate NOON states may also be generated by considering two-mode squeezed vacuum states as inputs~\cite{QOKnight}. Recent advances in the generation of high-intensity squeezed states of light have seen various applications, including the detection of gravitational waves~\cite{meylahn2022squeezed}. The intensity of nonclassical states may also be increased by displacing a squeezed state~\cite{thekkadath2022measuring}.

To increase the gravitational phase signal, one may also consider using fibre loops in the split branches of the interferometers. Along a single loop, the light pulses cross one another in opposite directions twice. In the first approximation, we can generalise this to $l$ loops by simply adding the actions, which amounts to multiplying the phase by a factor $2l$ for $l$ loops, where the length $D$ for the single-shot experiment would correspond to the length of half a loop. In all due rigour, one should carry out the derivation of the metric perturbations in the loop geometry where the light beams are not always counter-propagating in a fixed direction, one also needs to take into account the stress-energy contribution of light propagating in a medium as laid out in Ref.~\cite{Scully1979}. This is by no means a straightforward analysis, as adding fibres and having the light follow a longer optical path also comes with drawbacks, such as pulse broadening, power loss, thermal noise, birefringence etc.

Another drawback, that is not specific to the photonic GME protocol, is the presence of background noise. Although the photonic variant is devoid of additional unwanted known interactions, there remains background noise from the surrounding mass distribution. One way to offset the noise is to consider a background with a quasi-static mass density distribution. The experiment can be run first by considering co-propagating light pulses. In vacuum, no phase from the light-light coupling should be induced, and any signal will be due to noise. We can then compare this to the signal obtained with counter-propagating pulses and then conclude whether the entanglement generation was indeed between the light pulses.

Using a bosonic source comes with another advantage compared to using fermionic matter: the separation between the closest branches of the light beams may be tuned to be very small, even completely superimposed. This is conditioned to interference and diffraction being accounted for, and the experimental ability to distinguish between the pulses at close distances with the inclusion of diffraction effects~\cite{Schneiter_2018}.

In analogy to the generalisation to arbitrary geometries and qudits for the massive GME protocol~\cite{Tilly2021,Schut_2022}, one may also consider an array of light beams to boost the gravitational phase induction. In this manner, it may be interesting to examine whether there is any setup that entangles faster than the one studied in this work.

The photonic implementation may open up a new window into investigating gravity mediated entanglement, and using photons, as opposed to massive objects in a near-future experiment could also shed light on other theoretical aspects and features of gravity that cannot be witnessed using the latter.

\section*{Methods}

As we have seen, obtaining the entangling phase through the path integral formalism under the stationary phase approximation requires the evaluation of the classical action, an expression of which was shown in Eq.~\eqref{eq:action}. We recapitulate the assumptions of the model and show a step-by-step derivation of how the result is arrived at. By definition, the action is expressed as \begin{equation*}
S^{\sigma}_{\mathcal{F}}=\frac{1}{4}\int d^4 x h_{\mu\nu}T^{\mu\nu}.\end{equation*}
We use a pulse model for which the only non-zero stress-energy tensor component is of the form $Au(ct\pm z)\delta(x)\delta(y)$, where $A$ is the effective area of each pulse in the $xy$ plane, orthogonal to the propagation direction. Neglecting interference effects, which is valid for two sufficiently separated pulses, the total stress-energy tensor is then given by Eq.~\eqref{eq:Tfor2pulses}. Furthermore, under the weak-field limit (linearised gravity), one may simply add the two metric contributions sourced by each pulse. It follows that
\begin{equation*}
\begin{aligned}
S^{\sigma}_{\mathcal{F}}=A\sum_{a=1}^{2}
\int d^4 x h^p(\vec{x},t) 
u_a(z,t)\delta(x-x_a)\delta(y-y_a),
\end{aligned}
\end{equation*}
where $h^p$ is the metric perturbation for each pulse. Carrying out the integral over the $\delta$-distributions and further taking the polarisation of the pulses to be circular simplifies the expression to
\begin{equation*}
  S^{\sigma}_{\mathcal{F}}  =A u_0\sum_{a=1}^{2} \int dz dt h^p(x_a,y_a,z,t).
\end{equation*}
Hence, one has the explicit sum of two single pulse contributions
\begin{equation*}
    \begin{aligned}
    S^{\sigma}_{\mathcal{F}}&=A u_0\int dz dt\left( h^p(x_1,y_1,z,t)+h^p(x_2,y_2,z,t)\right)\\
&=A u_0\int_{z=-\infty}^{\infty}dz\int_{t=0}^{\tau}dt \left( h^p_0[0,0,z,t]+h^p_0[x_1-x_2,0,-z+D,t]+h^p_0[x_2-x_1,0,z,t]+h^p_0[0,0,-z+D,t]\right).
    \end{aligned}
\end{equation*}
Thus, we have broken down the task into four evaluations of circularly polarised single pulse metric perturbations. Each term can be expressed through Eq.~\eqref{eq:hp_expr}, yielding Eq.~\eqref{eq:action}, as announced.

\section*{Acknowledgements}
We are grateful to Dennis R{\"a}tzel, Sougato Bose, Yue Ma and Anupam Mazumdar for insightful discussions. We acknowledge financial support from the UK EPSRC through EP/T00097X/1 and EP/L016524/1, the UK STFC through ST/W006553/1 and the Korea Institute of Science and Technology through their Open Innovation programme.

\section*{Author contributions statement}
SA and HC wrote the main manuscript text, produced the numerics and prepared the figures, HC and MSK provided scientific guidance. All authors reviewed the manuscript.

\section*{Data availability statement}
The data that support the findings of this study are available from the corresponding authors upon reasonable request.

\bibliography{bibliography}

\appendix

\end{document}